\documentclass[%
 reprint,
 superscriptaddress,
 amsmath,amssymb,
 aps,
prb,
]{revtex4-2}

\usepackage{graphicx}
\usepackage{subfigure}
\usepackage{float}
\usepackage{dcolumn}
\usepackage{bm}
\usepackage{rotating}
\usepackage{color}
\usepackage{braket}
\usepackage[colorlinks,citecolor=blue,urlcolor=blue,linkcolor=blue]{hyperref}

\begin{document}

\preprint{APS/123-QED}

\title{Topological phases of extended Su-Schrieffer-Heeger-Hubbard model}

\author{Pei-Jie Chang}
    \affiliation{State Key Laboratory of Low-Dimensional Quantum Physics and Department of Physics, Tsinghua University, Beijing 100084, China}
    
\author{Jinghui Pi}
    \email{pijh14@gmail.com}
    \affiliation{State Key Laboratory of Low-Dimensional Quantum Physics and Department of Physics, Tsinghua University, Beijing 100084, China}
    
\author{Muxi Zheng}
    \affiliation{State Key Laboratory of Low-Dimensional Quantum Physics and Department of Physics, Tsinghua University, Beijing 100084, China}

\author{Yu-Ting Lei}
    \affiliation{State Key Laboratory of Low-Dimensional Quantum Physics and Department of Physics, Tsinghua University, Beijing 100084, China}

\author{Xingbo Pan}
    \affiliation{State Key Laboratory of Low-Dimensional Quantum Physics and Department of Physics, Tsinghua University, Beijing 100084, China}

\author{Dong Ruan}
    \affiliation{State Key Laboratory of Low-Dimensional Quantum Physics and Department of Physics, Tsinghua University, Beijing 100084, China}
    \affiliation{Frontier Science Center for Quantum Information, Tsinghua University, Beijing 100084, China}

\author{Gui-Lu Long}
    \email{gllong@tsinghua.edu.cn}
    \affiliation{State Key Laboratory of Low-Dimensional Quantum Physics and Department of Physics, Tsinghua University, Beijing 100084, China}
    \affiliation{Beijing Academy of Quantum Information Sciences, Beijing 100193, China}
    \affiliation{Frontier Science Center for Quantum Information, Tsinghua University, Beijing 100084, China}
    \affiliation{Beijing National Research Center for Information Science and Technology, Beijing 100084, China}

\begin{abstract}
Despite extensive studies on the one-dimensional Su-Schrieffer-Heeger-Hubbard (SSHH) model, the variant incorporating next-nearest neighbour hopping remains largely unexplored. Here, we investigate the ground-state properties of this extended SSHH model using the constrained-path auxiliary-field quantum Monte Carlo (CP-AFQMC) method. We show that this model exhibits rich topological phases, characterized by robust edge states against interaction. We quantify the properties of these edge states by analyzing spin correlation and second-order Rényi entanglement entropy. The system exhibits long-range spin correlation and near-zero Rényi entropy at half-filling. Besides, there is a long-range anti-ferromagnetic order at quarter-filling. Interestingly, an external magnetic field disrupts this long-range anti-ferromagnetic order, restoring long-range spin correlation and near-zero Rényi entropy. Furthermore, our work provides a paradigm studying topological properties in large interacting systems via the CP-AFQMC algorithm. 
\end{abstract}
\maketitle


\section{Introduction}

Topological phases of matter are a central topic in modern condensed matter physics due to their rich phenomenology and possible diverse applicability \cite{RevModPhys.82.3045,RevModPhys.83.1057,RevModPhys.88.035005,RevModPhys.89.041004}.This includes applications in topological spintronics \cite{nagaosa2013topological}, quantum  metrology \cite{pekola2013single}, and quantum computation \cite{beenakker2013search, li2014topological}. Within the independent electron approximation, topological matter can be characterized by the topological band theory \cite{PhysRevLett.98.106803, hsieh2008topological, zhang2009topological,RevModPhys.88.021004}. Specifically, the degeneracy of zero-energy edge modes is directly associated with the topology of bulk band through the bulk-edge correspondence \cite{PhysRevLett.89.077002,PhysRevB.74.045125}. There are various experimental platforms to explore the topological states of matter, such as ultracold atoms \cite{atala2013direct, cooper2019topological}, artificially engineered solid systems including graphene \cite{rizzo2018topological, pesin2012spintronics}, arrays of carbon monoxide molecules \cite{gomes2012designer, kempkes2019robust}, and exciton-polariton systems \cite{klembt2018exciton, su2021optical}. 

However, in many experiments setups, the interaction between electrons dominates over the hopping amplitude \cite{PhysRevB.96.245406, dusko2018adequacy, hu2023two}, rendering the independent-electron approximation invalid. Topological descriptions of strongly correlated systems are an emerging area of research with significant potential \cite{rachel2018interacting,junemann2017exploring, nawa2019triplon}. In interacting systems, the bulk-edge correspondence is modified as the relation between bulk topological property and the many-body, ground-state degeneracy on the edge \cite{PhysRevLett.109.096403}. Researchers have also developed a range of metrics to evaluate characterize topological edge states, including the Green function method \cite{manmana2012topological, wagner2023mott, wagner2023edge}, entanglement entropy between topological edges and bulk states \cite{jiang2012identifying, wang2015detecting}, and topological phase transition induced by magnetic flux changes \cite{xiao2010berry}. 

Su-Schrieffer-Heeger-Hubbard (SSHH) model with the nearest neighbour hopping has been extensively studied. However, the one with next-nearest neighbour hopping remains largely unexplored. Here, we discuss the ground-state properties of a one-dimensional SSHH model with next-nearest neighbour hopping. In the absence of Hubbard interaction, this model can be well described by the single-particle theory and exhibits four distinct topological phases ($\mathcal{W}=-1,0,1,2$) without breaking chiral symmetry \cite{PhysRevB.99.035146, PhysRevB.109.035114}. According to the bulk-edge correspondence, these topological phases are characterized by the degeneracy of zero-energy edge modes. However, in the presence of onsite Hubbard interaction, we should consider the relation between bulk topological property and the many-body ground-state. We evaluate long-range spin correlation and the second-order Rényi entropy of the ground state, and determine the topological state of the system by measuring the correlation between the boundary and bulk. The presence of Coulomb interaction leads to a Mott gap at half-filling and dimerization at quarter-filling \cite{le2020topological}. To mitigate the influence of finite site effects on our results, we employ the constrained-path auxiliary-field quantum Monte Carlo (CP-AFQMC) algorithm. Furthermore, the applicability of constrained-path method extends beyond the half-filling case \cite{wang2015detecting} where there is no sign problem. Our analysis reveals that the system exhibits long-range spin correlation and near-zero Rényi entropy at half-filling, indicative of a topological order state. Furthermore, a long-range antiferromagnetic order emerges at quarter-filling. The application of an external magnetic field disrupts this antiferromagnetic order, leading to a restoration of long-range spin correlation and near-zero Rényi entropy. 
 

The rest of the paper is organized as follows: In Sec. \ref{se2}, we simulate the SSH model incorporating next-nearest neighbour hopping and briefly review the exact diagonalization and CP-AFQMC methods. In Sec. \ref{se3}, we discuss the exist of long-range spin correlation and Rényi entanglement entropy at different filling. A summary is given in Sec. \ref{se4}. In the Appendix, we give result based on exact diagonalization as benchmark and result of CP-AFQMC with different parameters. 


\section{MODEL AND METHOD}\label{se2}

\begin{figure*}
     \centering
     \includegraphics[width=\textwidth]{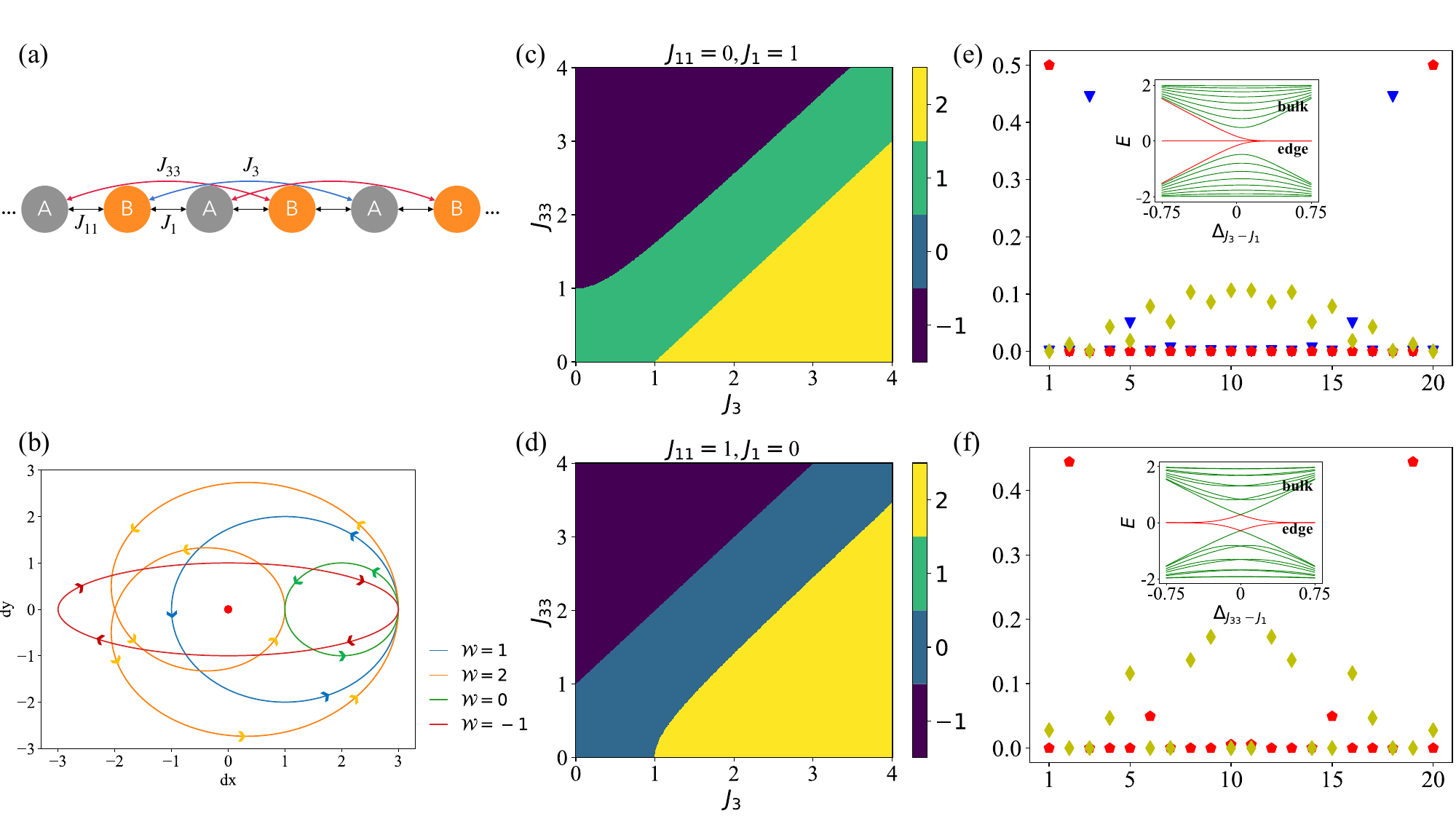}
     \caption{Su-Schrieffer-Heeger model with next-nearest neighbour hopping. (a) Schematic diagram of hopping term. (b) Schematic diagram of wave function rotation in pseudo-spin space. The solid point is the topological singular point and $\mathcal{W}$ is corresponding winding number. Topological phase diagram by calculating winding number with (c) $J_{11}=0,J_1=1$, (d) $J_{11}=1,J_1=0$ under periodic boundary condition. (e), (f) Under open boundary condition and along x(y) axis in (c), particle distribution of a typical bulk state (diamond) and edge state (five-pointed star and inverted triangle) for $\Delta_{J_3(J_{33})-J_1}=0.75$. We define that $\Delta_{J_3-J_1}=(J_3-J_1)/2$, $\Delta_{J_{33}-J_1}=(J_{33}-J_1)/2$ and $\Delta_{J_{33}-J_{11}}=(J_{33}-J_{11})/2$. Inset: Eigen-energies calculated for a chain with $N=20$ sites. The gapless band represents the topological edge state. }
     \label{fig1}
\end{figure*}

\subsection{Next-nearest neighbour hopping SSHH model}\label{2a}
Let's consider a SSHH model with next-nearest neighbour hopping. The Hamiltonian is given by
\begin{equation}
    H=H_0+H_U,
    \label{model}
\end{equation}
with hopping term
\begin{equation}
    \begin{aligned}
        H_{0}= &  -\sum_{i,\sigma}(J_{11}\hat{c}_{i,\sigma}^{A\dagger}\hat
        {c}_{i,\sigma}^{B}+J_{1}\hat{c}_{i,\sigma}^{B\dagger}\hat{c}_{i+1,\sigma}
        ^{A}+J_{33}\hat{c}_{i,\sigma}^{A\dagger}\hat{c}_{i+1,\sigma}^{B}\\
        &  +J_{3}\hat{c}_{i\sigma}^{B\dagger}\hat{c}_{i+2,\sigma}^{A}+h.c)\newline,
    \end{aligned}
\end{equation}
and onsite interaction term  
\begin{equation}
    H_U = \sum_{i,s} U\hat{n}^{s}_{i\uparrow}\hat{n}^{s}_{i\downarrow},
\end{equation}
where $\hat{c}_{i\sigma}^{s\dagger}$ is the fermion creation operator of the $i$-th site with sub-lattice label $s$ and spin label $\sigma$. The parameter $J_1$ represents inter-cell hopping and $J_{11}$ represents intra-cell hopping which are similar to SSH model,  $J_3$ and $J_{33}$ are the next-nearest neighbour hopping between A and B as shown in Fig.~\ref{fig1}(a). $U$ is the on-site interaction strength with particle number operator $n^s_{i\sigma}=\hat{c}_{i\sigma}^{s\dagger}\hat{c}_{i\sigma}^{s}$.

We first discuss the topological phases of the SSH model with next-nearest neighbour hopping. The Bloch Hamiltonian of this model is 
\begin{equation}
    \mathcal{H}(k)=d_x(k)\sigma_x+d_y(k)\sigma_y,
    \label{4}
\end{equation}
which is obtained by the Fourier transform of $H_0$ under the periodic boundary condition, where $d_x=-J_{11}-(J_{1} + J_{33})\cos k -J_{3}\cos 2k$ and $d_y=-J_{11}-(J_{1} -J_{33})\sin k -J_{3}\sin 2k$, $\sigma_{x(y)}$ are the Pauli matrices. 

The model (\ref{4}) exhibits time-reversal symmetry $\mathcal{T}$, defined as $\mathcal{T}\mathcal{H}(k) \mathcal{T}^{-1} = \mathcal{H}(-k)$. This symmetry satisfies $ \mathcal{T}^2=1$ and takes the form $\mathcal{T}=K$, where $K$ is the complex conjugation operator. Furthermore, the model preserves chiral symmetry $\mathcal{C}=\sigma_z$, which anticommutes with the Hamiltonian, $\sigma_z \mathcal{H} = -\mathcal{H} \sigma_z $. Consequently, each eigenstate $\ket{\psi}$ with energy $E$ possesses a corresponding partner state $\sigma_z \ket{\psi} $ with energy $-E$. In addition, this model has particle-hole symmetry $\mathcal{S}$, which is the combination of time-reversal symmetry and chiral symmetry, $\mathcal{S}= \mathcal{T} \mathcal{C} $. This symmetry also anticommutes with the Hamiltonian, $\mathcal{S} \mathcal{H} = -\mathcal{H} \mathcal{S} $. Thus, according to the tenfold way method, this model belong the BDI topological class as it possesses the above three symmetry \cite{Ryu_2010,RevModPhys.88.035005}.

To characterize the topological property of the interaction free model (\ref{4}), we can define the winding number $\mathcal{W}$ as follow \begin{equation}
    \mathcal{W}=\frac{1}{2 \pi} \int_{\mathrm{FBZ}} \frac{d_x \partial_k d_y-d_y \partial_k d_x}{d_x^2+d_y^2} d k,
\end{equation}where the integral is a close loop with $k$ varying in the first Brillouin zone (FBZ), namely $ k\in\lbrack0,2\pi) $ as we set unit cell length $a=1$. We plot the typical $\mathcal{W}$ in the parameter space spanned by $d_x$ and $d_y$ in Fig.~\ref{fig1}(b). This winding number can be associated to the Zak phase \cite{PhysRevLett.62.2747, xiao2010berry}\begin{equation}
\gamma=i\int_{\mathrm{FBZ}}\left\langle \psi|\partial_{k}|\psi\right\rangle dk
\end{equation}via $\gamma=\pi \mathcal{W}$, where $\left\vert \psi\right\rangle =\frac{1}{\sqrt{2}}\left(  \frac{d_{x}-id_{y}}{\sqrt{d_{x}^{2}+d_{y}^{2}}},-1\right)^{T}$ is the normalized Bloch eigenstate with eigenvalue $\sqrt{d_{x}^{2}+d_{y}^{2}}$. If the next-nearest neighbour hopping $J_3=J_{33}=0$, the model is the standard SSH model, which have winding number $\mathcal{W}=0$ or $1$, corresponding to topological trivial phase or topological nontrivial phase, respectively. When $J_3$ and $J_{33}$ is non-zero, the phase diagram undergoes a modification, allowing $\mathcal{W}$ to assume the additional values of $-1$ and $2$, as shown in Fig.~\ref{fig1}(c) and~\ref{fig1}(d). 

According to the bulk-edge correspondence principle, the winding number $\mathcal{W}$ defined by the bulk Hamiltonian (\ref{4}) is directly related to the topological edge states  under open boundary condition. Specifically, at $\mathcal{W}=0$, the system lacks topological edge states. Conversely, for $\mathcal{W}=\pm1$, two topological edge states with zero energy emerge at different sites. When $\mathcal{W}=2$, the number of zero energy topological edge states increases to four. We show the typical topological edge states at $1_{st}, 3_{rd}, (N-2)_{th}, N_{th}$ sites with $\mathcal{W}=2$ in Fig.~\ref{fig1}(e), and topological edge states at $2_{nd},(N-1)_{th}$ sites with $\mathcal{W}=-1$ in Fig.~\ref{fig1}(f). 

In the presence of finite on-site Hubbard interaction ($U \neq 0$), the analytical solution of model (\ref{model}) via the Bethe ansatz becomes intractable in general \cite{bethe1997theory}. In the flowing part, we investigate the effect of $U$ using numerical simulation techniques. For clarity, the subsequent two sections provide a concise overview of the two numerical methods employed in this work. 

\subsection{Exact diagonalization}
Exact diagonalization (ED) serves as a benchmark for solving quantum mechanics problems numerically. In theory, it provides a path to solve any problem given sufficient computational resources. However, condensed matter lattice models often present exponentially difficult complexities. For instance, the fermion Hubbard model scales as $4^N$, where $N$ represents the number of lattice sites. This rapid growth renders simulations of large-scale models computationally prohibitive. Despite these limitations, ED's accuracy and reliability make it valuable for obtaining reliable results in small-scale quantum simulations. We consider an open-boundary, one-dimensional chain with $N=8$ sites. Conserved particle number guarantees a $U(1)$ symmetry in the system. This symmetry allows for block-diagonalization of the Hamiltonian, facilitating efficient diagonalization using the Lanczos algorithm. However, ED scales exponentially, restricting its application to small systems and inherently limiting studies of finite-size effects. To overcome this limitation and access the thermodynamic limit, we employ the constrained-path auxiliary-field quantum Monte Carlo method, detailed in the following section.

\subsection{Constrained-path auxiliary-field quantum Monte Carlo}
In this section, we give a brief introduction to constrained-path auxiliary-field quantum Monte Carlo (CP-AFQMC), which proves to be a powerful tool for solving the ground state problem in many-body systems \cite{zhang1995constrained, zhang1997constrained, zhang201315, nguyen2014cpmc, PhysRevB.99.045108}. Analogous to the power method for finding the dominant eigenvalue and eigenvector of a matrix  \cite{panza2018application}, iterative application of the imaginary time evolution operator to an initial state $\ket{\psi_0}$ recovers the ground state  $|\psi_g\rangle$, provided the overlap is nonzero,
\begin{equation}
    |\psi_g\rangle\propto \mathop{lim}\limits_{\beta\to\infty} e^{-\beta H}|\psi_0\rangle,\ if\ \langle\psi_g|\psi_0\rangle\neq0.
    \label{img}
\end{equation}

To simplify numerical simulations, we assume the initial state $\ket{\psi_0}$ to be a single Slater determinant, as employed in Ref.~\cite{nguyen2014cpmc}. Besides, as the edge states come in degenerate pairs, constructing an appropriate $|\psi_0\rangle$ necessitates a suitable linear superposition of these states. For instance, if the zeros energy edge state of single particle states are 
$\ket{\phi_1}=(1,0,\cdots,0)^T,\ \ket{\phi_2}=(0,\cdots,0,1)^T,$
then the initial state $|\psi_0\rangle$ can be chosen to $\ket{\psi_1}=(\ket{\phi_1}+ \ket{\phi_2})/\sqrt{2}$ or $\ket{\psi_2}=(\ket{\phi_1}-\ket{\phi_2})/\sqrt{2}$.  

Using Trotter decomposition, we can separate the imaginary time evolution process into single-body and multi-body contributions
\begin{equation}
        e^{-\beta H}=(e^{-\tau H})^n=(e^{-\frac12\tau H_0}e^{-\tau H_U}e^{-\frac12\tau H_0})^n+O(\tau^2),
\end{equation}
where $\beta=\tau n$ is the length of imaginary time and $\tau$ is one step of imaginary time. In the calculation of AFQMC, we can simply calculate that a single body operator on the Slater determinant, and the result can still be expressed by another Slater determinant. As two-body terms are incompatible with the Slater determinant formalism, then the Hubbard-Stratonovich transformation is employed. This approach decouples the two-body Hamiltonian into a superposition of single-body Hamiltonians by introducing an auxiliary field \cite{PhysRevB.28.4059}. Thus far, all operators have been represented as single-body operators, facilitating efficient computations within Slater determinants. In AFQMC, the wave function is represented as a linear combination (where is a single) of Slater determinants (walkers). Eq.~(\ref{img}) is then represented as random walks in the Slater determinant space by sampling the auxiliary fields.

A significant challenge in simulating quantum systems with the AFQMC method is the sign problem. This arises from negative weights assigned to certain configurations during the simulation, leading to a substantial increase in variance and consequently unreliable results. To address this challenge, CP-AFQMC utilizes a trial wave function, denoted as $|\psi_T\rangle$. A key point is that walkers have a negative overlap with trial wave function are excluded, while those with significant overlap are replicated for propagation. Following extended imaginary time evolution, the converged ground state energy and its associated wave function can be determined. A limitation of this method arises from the systematic errors introduced into the simulation results by the selection of $|\psi_T\rangle$. Optimizing $|\psi_T\rangle$ is an active research area within the CP-AFQMC framework \cite{qin2023self, PhysRevA.100.023621}. Here, we choose the initial state $\ket{\psi_0}$ as the trial wave function. Our primary focus is the characterization of edge states. Consequently, their existence or absence plays a crucial role in our simulations. This necessitates the implementation of a suitable linear superposition of $\ket{\phi_1}$ and $\ket{\phi_2}$, as previously established. 

Physical quantities can be obtained using the equal-time Green's function, formulated as the mixed estimator, 
\begin{equation}
    G_{ij}=\langle\hat{c}_{j}^{\dagger}\hat{c}_{i}\rangle=\frac{\sum_kw_k\langle\psi_T|\hat{c}_j^{\dagger}\hat{c}_i|\psi_k\rangle}{\sum_kw_k\langle\psi_T|\psi_k\rangle}, 
    \label{green}
\end{equation}
where $|\psi_k\rangle$ is the $k_{th}$ walker, $\omega_k$ is the corresponding weight of the walker, and $|\psi_T\rangle$ is the trail wave function where same as $\ket{\psi_0}$. Specifically, by employing the Wick's theorem \cite{PhysRev.80.268}, physical observations like spin correlation and Rényi entanglement entropy can be calculated via the equal-time Green's function. 

\section{Result}\label{se3}

\begin{figure}
    \centering
    \includegraphics[width=0.4\textwidth]{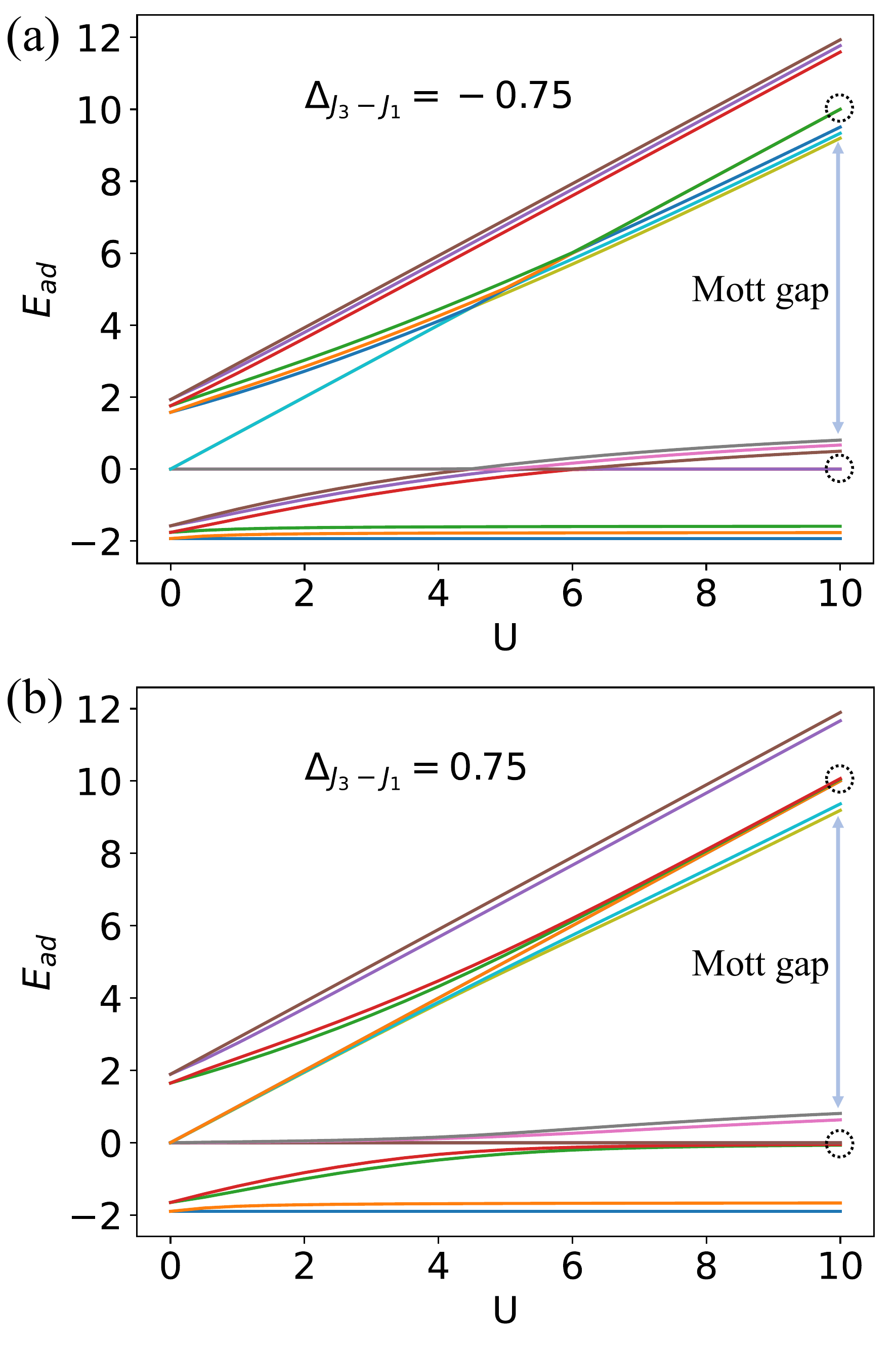}
    \caption{The addition energy spectrum of the next-nearest neighbour hopping Su-Schrieffer-Heeger-Hubbard model for a $N=8$ sites chain as a function of Hubbard $U$ with (a) $\Delta_{J_3-J_1}=-0.75$, (b) $\Delta_{J_3-J_1}=0.75$. The circled areas are the gapless states, and the position circled in b is two lines, it is difficult to distinguish because the values are very close.}
    \label{fig2}
\end{figure}

\begin{figure*}
     \centering
     \includegraphics[width=\textwidth]{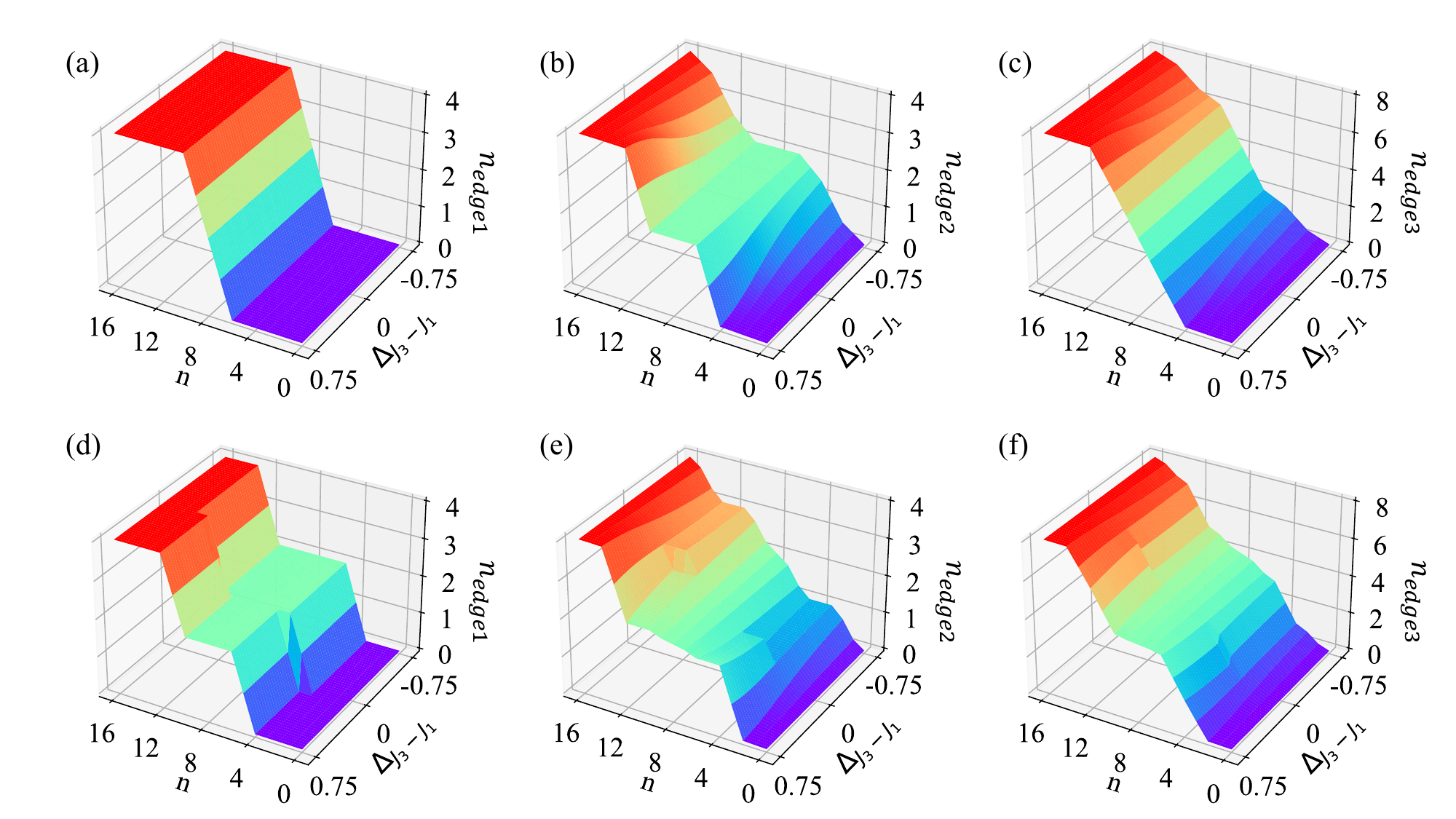}
     \caption{Population number of the edge states for $N=8$ sites chain with $J_{11}=J_{33}=0$, and $n$ is the number of electrons. (a)-(c) for $U=0$ and (d)-(f) for $U=10$. According to Fig.~\ref{fig1}(e), the edge states appear at the first boundary ($1_{st}$ and $N_{th}$) sites or the third-boundary ($3_{rd}$ and $(N-2)_{th}$) sites. (a), (d) $n_{edge1}$ is the sum of electron number on the $1_{st}$ and $N_{th}$ sites; (b), (e) $n_{edge2}$ is the sum of electron number on the $3_{rd}$ and $(N-2)_{th}$ sites; (c), (f) $n_{edge3}$ is the sum of electron number on above four sites. In the non-trivial phases without Hubbard $U$, the edge population increases sharply at half-filling, while with Hubbard $U$ the edge population increases sharply at quarter-filling and there-quarter-filling. And in the trivial phase, the edge population increases gradually. There are some wrinkles on the surface in (d)-(f), which is due to the finite-size effects. }
     \label{fig3}
\end{figure*}

\begin{figure*}
     \centering
     \includegraphics[width=\textwidth]{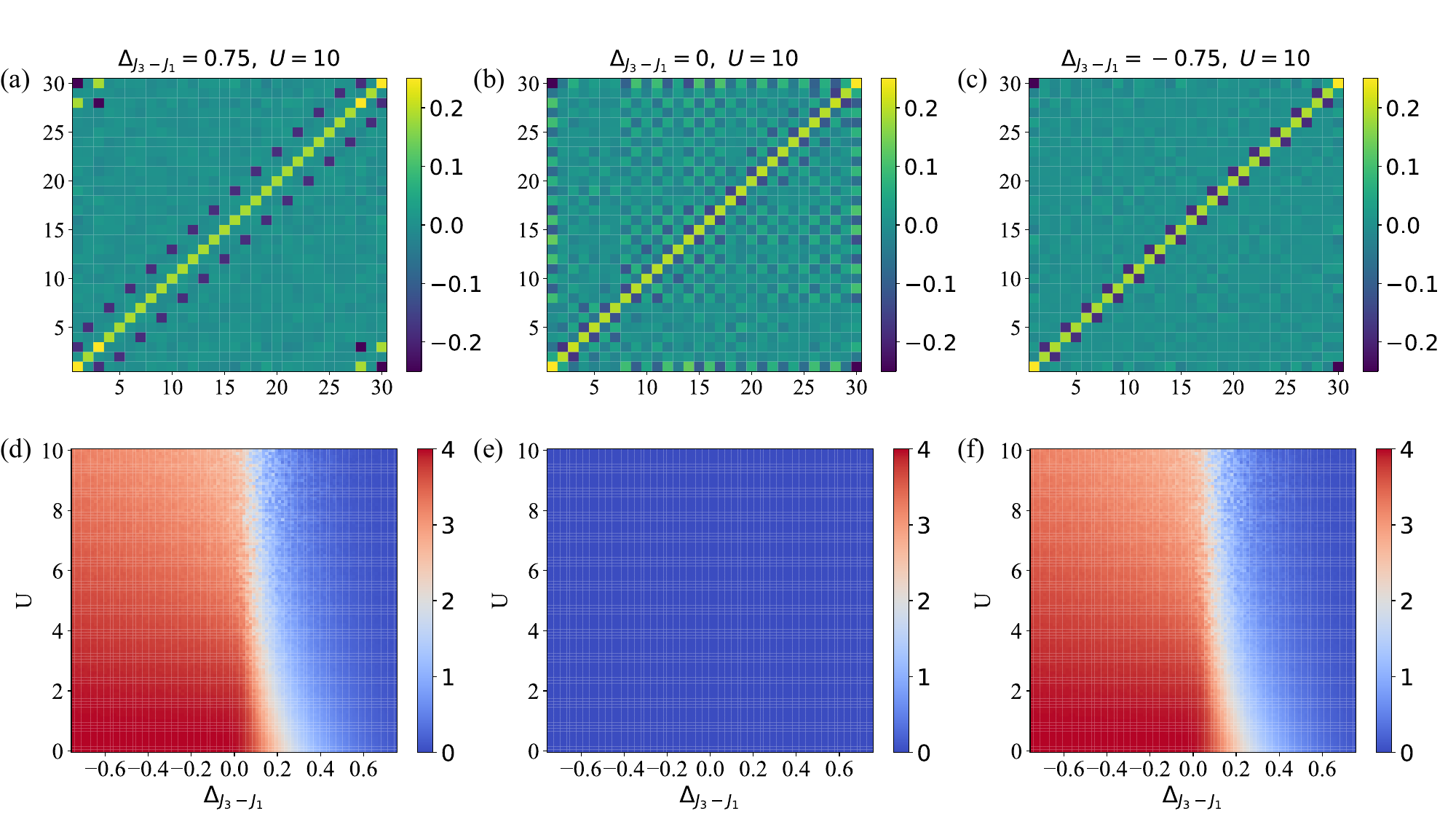}
     \caption{CP-AFQMC results of next-nearest neighbour hopping SSHH model with OBC at the half-filling with $J_{11}=J_{33}=0$. (a) Spin correlation $\langle S_{z,j}S_{z,k}\rangle$ with $\Delta_{J_3-J_1}=0.75$, topological edge states at [$1_{st}, 3_{rd}, (N-2)_{th}, N_{th}$] sites. (b) Same as (a) with $\Delta_{J_3-J_1}=0$ and there is a long-range anti-ferromagnetic order. (c) Same as (a) with $\Delta_{J_3-J_1}=-0.75$, and topological edge states at [$1_{st}, N_{th}$] sites. (d) The entanglement Rényi entropy between subsystem [$3_{rd}, (N-2)_{th}$] sites and the remaining sites; (e) Same as (d) but subsystem [$1_{st}, N_{th}$] sites and (f) with subsystem [$1_{st}, 3_{rd}, (N-2)_{th}, N_{th}$] sites. In (d)-(f), the position where the transition occurs is closer to $\Delta _ {J _ 3 - J _ 1}=0$ than ED in Appendix \ref{appendixC}, which means that the finite-size effects have less impact of the conclusion. }
     \label{fig4}
\end{figure*}

\begin{figure*}
     \centering
     \includegraphics[width=\textwidth]{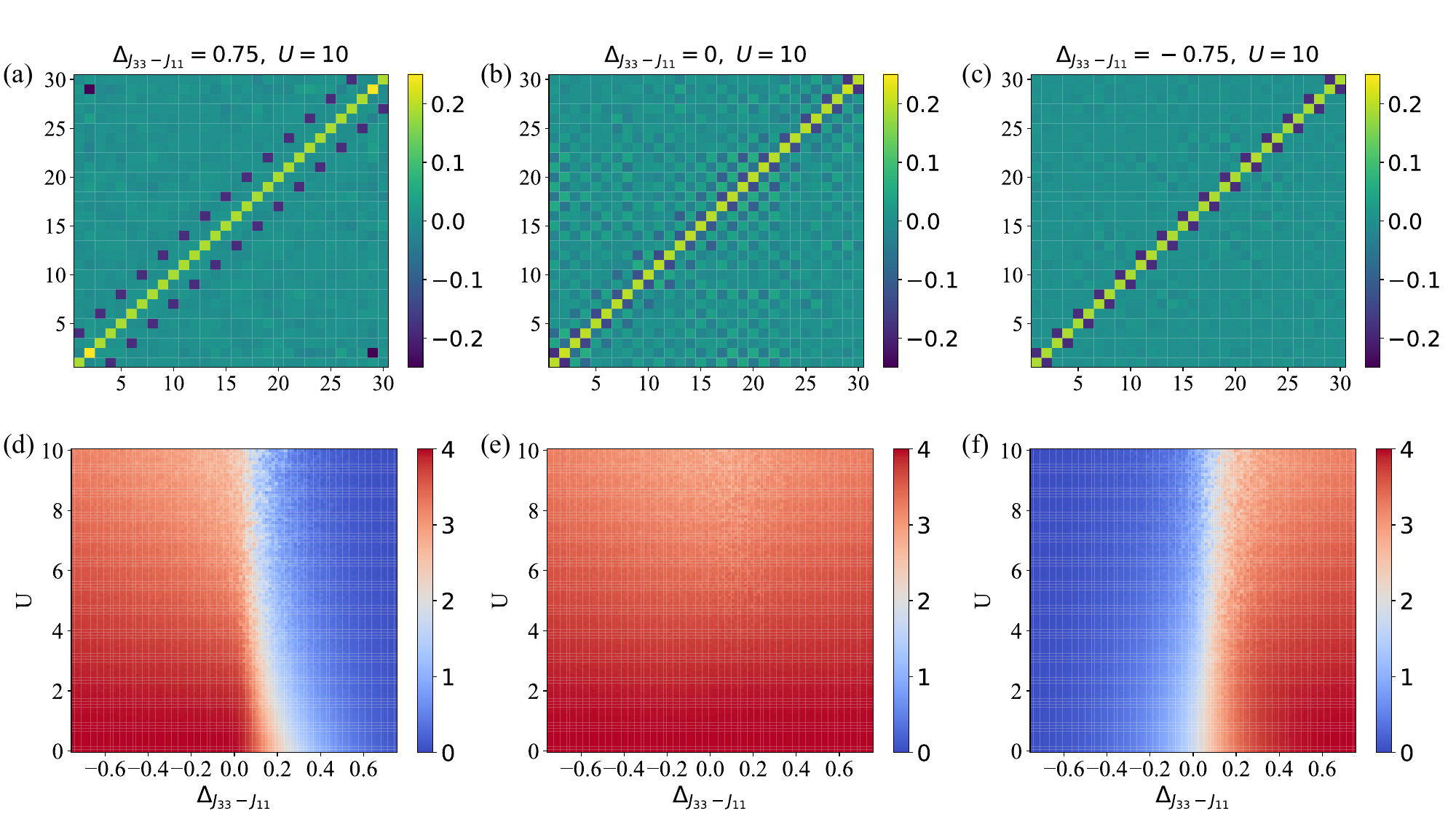}
     \caption{CP-AFQMC results same as Fig~\ref{fig4} but with $J_{1}=J_{3}=0$. (a) Spin correlation $\langle S_{z,j}S_{z,k}\rangle$ with $\Delta_{J_{33}-J_{11}}=0.75$, topological edge states at [$2_{nd}, (N-1)_{th}$] sites. (b) Same as (a) with $\Delta_{J_{33}-J_{11}}=0$ and there is a long-range anti-ferromagnetic order. (c) Same as (a) with $\Delta_{J_{33}-J_{11}}=-0.75$, and there are bulk dimers without topological edge states as shown in Fig~\ref{fig1}(d) $\mathcal{W}=0$. (d) The entanglement Rényi entropy between subsystem [$2_{nd}, (N-1)_{th}$] sites and the remaining sites; (e) Same as (d) but subsystem [$1_{st}, N_{th}$] sites and (f) with subsystem [$1_{st}, 2_{nd}, (N-1)_{th}, N_{th}$] sites. Due to the formation of dimers, there will be a drop in Rényi entanglement entropy. }
     \label{fig5}
\end{figure*}

\begin{figure*}
     \centering
     \includegraphics[width=\textwidth]{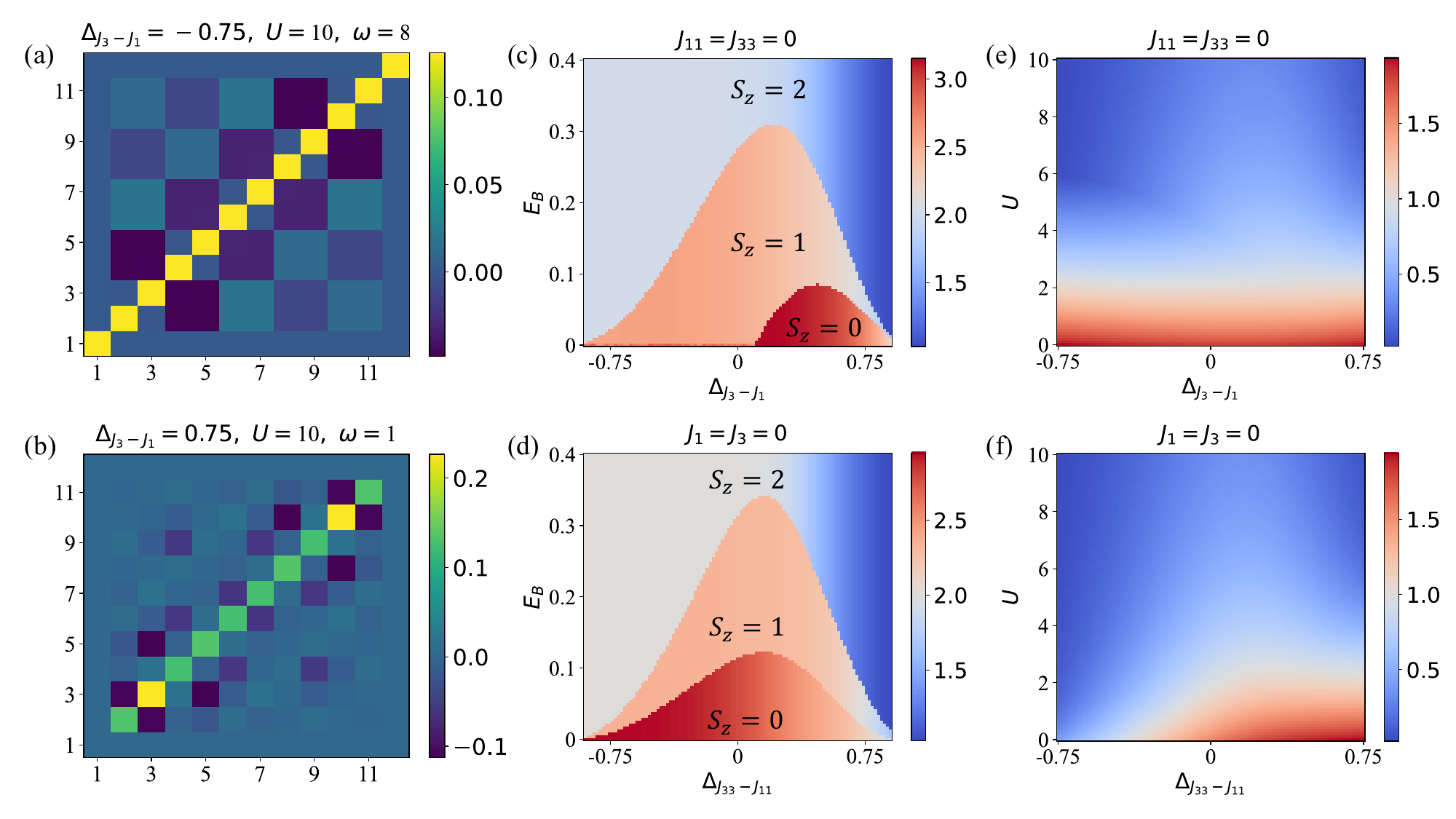}
     \caption{Exact diagonalization results of next-nearest neighbour hopping SSHH model with OBC at the quarter-filling with $J_{11}=J_{33}=0$ or $J_{1}=J_{3}=0$ for (b)(d)(f). (a) Spin correlation $\langle S_{z,j}S_{z,k}\rangle$ with $\Delta_{J_3-J_1}=-0.75,\ J_{11}=J_{33}=0$, topological edge states are [$1_{st}, N_{th}$] sites, and $\omega$ is degeneracy of ground state. (b) Same as (a) with $\Delta_{J_3-J_1}=0.75$ and topological edge states are [$1_{st}, 3_{rd}, (N-2)_{th}, N_{th}$] sites. (c) Rényi entanglement entropy between the edges [$3_{rd}, (N-2)_{th}$] sites and the bulk for various field strength and $\Delta_{J_3-J_1}$. And each contour can separate with different $S_z$, where $S_z=2$ means the maximally ferromagnetic ground state. (d) Same as (c) but with topological [$2_{nd}, (N-1)_{th}$] sites. (e) Critical field strength at various Hubbard $U$ and $\Delta_{J_3-J_1}$. (f) Same as (e) but changing with $\Delta_{J_{33}-J_{11}}$. If we do a $1D$ slice for $\Delta_{J_{3}-J_1}=0.75$ in (e) or $\Delta_{J_{33}-J_{11}}=0.75$ in (f), there will be a decrease in the critical field with increasing Hubbard $U$. (a) and (b) for a $1D$ chain with $N=12$, and (c)-(f) for a 1D chain with $N=8$. }
     \label{fig6}
\end{figure*}

\begin{figure*}
     \centering
     \includegraphics[width=\textwidth]{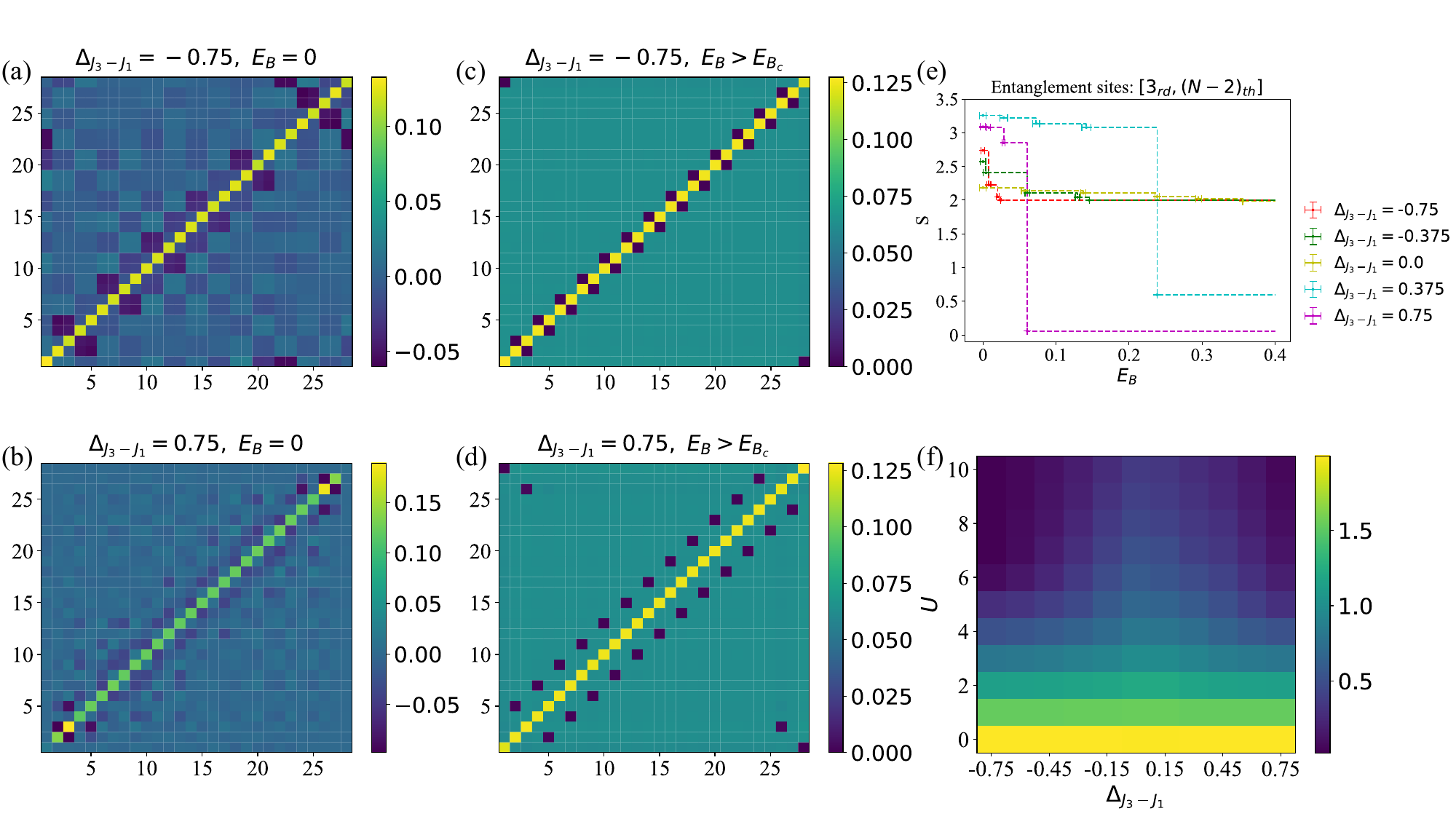}
     \caption{CP-AFQMC results of next-nearest neighbour hopping SSHH model with OBC at the quarter-filling with $J_{11}=J_{33}=0$, and $U=10$ for (a)-(e). (a) Spin correlation $\langle S_{z,j}S_{z,k}\rangle$ with $\Delta_{J_3-J_1}=-0.75$, topological edge states at [$1_{st}, N_{th}$] sites. (b) Same as (a) with $\Delta_{J_3-J_1}=0.75$ and topological edge states at [$1_{st}, 3_{rd}, (N-2)_{th}, N_{th}$] sites. (c) Spin correlation $\langle S_{z,j}S_{z,k}\rangle$ with $\Delta_{J_3-J_1}=-0.75$ above the critical magnetic field. (d) Same as (c) but $\Delta_{J_3-J_1}=0.75$. (e) Rényi entanglement entropy with changing external magnetic field and hopping amplitude difference $\Delta_{J_3-J_1}$. Each point has error bar in both x and y direction, and the dotted line is a schematic line connecting different $\Delta_{J_3-J_1}$. (f) Critical field strength at various of the on-site interaction and hopping amplitude difference and it can be seen that for a certain $\Delta _ {J _ 3 - J _ 1}$, the value of the critical magnetic field will decrease as $U$ increases. }
     \label{fig7}
\end{figure*}

To benchmark the CP-AFQMC calculations, we perform exact diagonalization on the chain with $N=8$ sites. Subsequently, CP-AFQMC simulations are carried out for a larger chain of $N=30$ sites. The system exhibits no sign problem at half-filling, allowing us to employ parameters of $\tau=0.05,\ N_{walker}=100,$ and $ \beta=n \tau=15.$ For systems away from half-filling, where a sign problem arises, we set $\tau=0.01,\ N_{walker}=2000,$ and $\beta=n\tau=64.$

\subsection{Edge population}

In the absence of Hubbard interaction ($U=0$), as discussed in Sec.~\ref{2a}, the chiral symmetry ensures that the eigenvalues appear in $(E,-E)$ pairs, leading to the emergence of topological edge states at the $E=0$. According to Fig.~\ref{fig1}(e) and~\ref{fig1}(f), while electrons progressively fill the bulk bands, they do not populate the edge states until the half-filling regime is reached. The occupation of the edge state commences, as illustrated in Fig.~\ref{fig3}(a). Under the parameters in Fig.~\ref{fig3} (where $J _ 1>J _ {11}$), the two outermost sites $1_{st}$ and $N_{th}$  always support edge states. Consequently, modifying the $\Delta _ {J _ 3 - J _ 1} $ will not affect the filling of this specific pair of edge states. 

The difference between nearest neighbour hopping and next-nearest hopping $\Delta _ {J _ 3 - J _ 1}$ affects the topological properties of $3_{rd}$ and $(N - 2)_{th}$ sites according to Fig.~\ref{fig1}(c). When $\Delta _ {J _ 3 - J _ 1}$ is negative , the $3_{rd}$ and $(N - 2)_{th}$ sites do not support edge states. Consequently, the electron population at these sites continuously varies across the entire filling range, mirroring the behavior of bulk states. Conversely, for positive $\Delta _ {J _ 3 - J _ 1}$, a pair of edge states appears at the $3_{rd}$ and $(N - 2)_{th}$ sites. This scenario, as illustrated in Fig.~\ref{fig3}(b), leads to a sharp change in the electron population at half-filling. By focusing on the combined behavior of these four potential edge state sites and considering the electron filling, we observe that only all four sites support topological edge state does a sharp population change occur at half-filling, as shown in Fig.~\ref{fig3}(c). 

In the strong interaction limit (large Hubbard $U$), the Coulomb repulsion between spin-up and spin-down electrons leads to the emergence of a Mott gap in the energy spectrum. This is reflected in the addition energy spectrum, $E_{ad}(n)=E(n)-E(n-1)$ where $n$ represents the number of electrons. As expected for the Hubbard model, $E_{ad}(n)$ separates the lower and upper bands, as illustrated in Fig.~\ref{fig2}(a) and \ref{fig2}(b). Notably, when Hubbard $U$ is sufficiently large, the edge states begin to be filled with electrons at quarter-filling. When considering different electron filling at this time, there will be a difference from no interaction, that is, electrons begin to occupy the edge state at quarter-filling as shown in Fig.~\ref{fig3}(d)-\ref{fig3}(f).  This discrepancy arises due to the inherent gapless nature of symmetry-protected topological edge states. When the system is quarter-filling with large Hubbard $U$, introducing an additional electron at any bulk site incurs a significant energy penalty due to the Hubbard $U$ interaction. This additional electron can occupy the edge sites without costing energy. 

There are some wrinkles on the surface in Fig.~\ref{fig3}(d)-\ref{fig3}(f). This is due to the fact that $3_{rd}$ and $(N-2)_{th}$ sites to support new topological edge states following a change in the topological number $\mathcal{W}$, which allows for an additional electron to occupy these two sites without costing additional energy. And filling the sites of $1_{st}$ and $N_{th}$ extends backward by one electron. As established in Fig.~\ref{fig1}(b), the system undergoes a topological phase transition at $J _ 3=J _ 1$, corresponding to $\Delta _ {J _ 3 - J _ 1}=0$. However, the wrinkles on the surface in Fig.~\ref{fig3}(d)-\ref{fig3}(f) deviate from zero. This is actually the result of the finite-size effects in the system.

\subsection{Bulk-edge correspondence}

The bulk-edge correspondence means that the topological properties of a system can be extracted by analyzing the characteristics of its edge states. The topological winding number, determined under periodic boundary conditions, captures the nature of the bulk states. Conversely, well-defined boundaries, enforced through open boundary conditions in simulations, are necessary for the emergence of edge states. This correspondence dictates that a change in the winding number induces the appearance of localized, gapless edge states that typically come in pairs. As illustrated in Fig.~\ref{fig2} and \ref{fig3}, regardless of the presence or absence of strong interaction, the edge state of the system is always occupied by electrons when it is half-filling. 

We choose $U=10,\ J_{11}=J_{33}=0$ for the investigation of the impact of $\Delta_{J_3 - J_1}$ on the system , as illustrated in Fig.~\ref{fig4}. In this configuration, $1_{st}$ and $N_{th}$ sites consistently support topological edge states. Additionally, for $\Delta_{J_3-J_1}>0$, $3 _ {rd}$ and $(N - 2) _ {th}$ sites also support topological non-trivial edge states. These edge states are decoupled from the bulk, and each pair of edge states existing in either a singlet state $\ket{S}=(\ket{\uparrow\downarrow}-\ket{\downarrow\uparrow})/\sqrt{2}$ or one of the three triplet states $\ket{T_0}=(\ket{\uparrow\downarrow}+\ket{\downarrow\uparrow})/\sqrt{2}$, $\ket{T_1}=\ket{\uparrow\uparrow}$ and $\ket{T_2}=\ket{\downarrow\downarrow}$. All there states are zero-energy and gapless. When $\Delta_{J_3 - J_1}<0$, the ground state of half-filling exhibits fourfold degeneracy, and for $\Delta_{J_3-J_1}>0$, the degeneracy of the ground state at half-filling increases to $16$, as confirmed by the exact diagonalization results presented in Appendix \ref{appendixC}. The existence of Hubbard $U$ reduces the ground state degeneracy. This arises due to the exclusion principle enforced by U,  forbidding the co-occupation of a single site by electrons with opposing spins.

As the edge states forming a decoupled single state or a triplet state, the topological properties of the bulk can be observed by simulating spin correlation between different sites. We define spin correlation $C_{j, k} = \langle S_{z,j}S_{z,k}\rangle$, where $S_{z,j}=\frac{1}{2}(\hat{n}_{j,\uparrow}-\hat{n}_{j,\downarrow})$ represents the z-component spin operator at site $j$. According to the equation (\ref{green}) and Wick's theorem \cite{PhysRev.80.268}, the calculation of the spin correlation function can be achieved through the two-point correlation Green function, as detailed in Appendix \ref{appendixB}.

As the $1_{st}$ and $N_{th}$ sites can always support topological edge states, the $1_{st}$ site only has spin correlation with $N_{th}$ site as shown in Fig.~\ref{fig4}(a)-\ref{fig4}(c). When $\Delta_{J_3 - J_1}>0$  [Fig.~\ref{fig4}(a)], a similar long-range spin correlation emerges between $3_{rd}$ site and $(N-2)_{th}$ site same as $1_{st}$ site and $N_{th}$ site. This is because the $3_{rd}$ and $(N-2)_{th}$ can also support topological non-trivial edge states. Conversely, when $3_{rd}$ and $(N-2)_{th}$ sites only support topological trivial states, long-range spin correlation are no longer observed, and a long range anti-ferromagnetic order is shown in Fig.~\ref{fig4}(b). This observation is consistent with Hubbard model's conclusion in the large $U$ limit, which can be attributed to second-order perturbation theory. 

Long-range correlation effect can be observed not only through spin correlation, but also by quantifying entanglement entropy. The paired appearance of topological edge states means the existence of near-zero entanglement. Since the positions of edge states are located at the two boundaries of the system, the formation of topological edge states can be determined by calculating the long-range entanglement entropy between the edge sites and bulk sites. 

Define the reduced density matrix of subsystem A as $\rho_A =Tr_B\ket{\Psi}\bra{\Psi}$, where A is $1_{st}$ and $N_{th}$ sites or $3_{rd}$ and $(N-2)_{th}$ sites. Rényi entropy can be calculated by the reduced density matrix of subsystem A as $S_n = -\frac{1}{n-1}\log_2[Tr(\rho_A^n)]$, where $n$ is a positive integer representing the order of the Renyi entropy. And the entanglement between subsystems A and B can be measured by Rényi entropy. In CP-AFQMC, the second-order Rényi entropy can be efficiently computed through the Green's function under different auxiliary fields \cite{PhysRevLett.111.130402, PhysRevB.89.125121}:
\begin{equation}
    \begin{aligned}S_2=&-\log\biggl[\sum_{\{s\},\{s^{\prime}\}}P_sP_{s^{\prime}}\{\mathrm{Det}(G_{s,A}G_{s^{\prime},A}\\
    &+(1-G_{s,A})(1-G_{s^{\prime},A}))\biggr]
    \end{aligned}
    \label{Rényi}
\end{equation}
where $G_s$ is equal-time Green matrix, and $G_s^{ij}=\langle \pmb{\hat{c}}^{\dagger}_j \pmb{\hat{c}}_i \rangle$ with $\pmb{\hat{c}}_i= \left(
\begin{matrix}
\hat{c}_{i\uparrow}, \hat{c}_{i\downarrow}
\end{matrix}
\right)$. 


The presence of topological edge states can be judged using the second-order Rényi entanglement entropy. When topological edge states exist, the entanglement between the sites where the topological edge states appear and the remaining system will be significantly reduced, as illustrated in Fig.~\ref{fig4}(d)-~\ref{fig4}(f). Since $1_{st}, N_{th}$ sites can always support topological edge states, leading to near-zero entanglement with the bulk, as depicted in Fig.~\ref{fig4}(e). Conversely, when $\Delta_{J_3-J_1}$ is changed until $3_{rd}, (N-2)_{th}$ can support a pair of topological edge states, the entanglement entropy also decrease close to zero. In all other configurations, the entanglement remains non-zero. Notably, the observed decrease in entanglement entropy does not occur precisely at $\Delta_{J_3-J_1}=0$ due to finite-size effects within the system. This effect will decrease as the system size increases. This observation aligns with the exact diagonalization results presented in the Appendix \ref{appendixC}. As larger system sizes calculated by CP-AFQMC, the position where the entanglement entropy decreases is closer to $\Delta_{J_3-J_1}=0$. 

Our calculations show that $J_{3}=J_1=0$ in Fig.~\ref{fig5}, and the conclusion obtained is consistent with Fig.~\ref{fig4}. When $\Delta_{J_{33}-J_{11}}<0$,\ $1_{st},\ 2_{nd},\ (N-1)_{th}$ and $\ N_{th}$ sites cannot support topological edge states. This is because $J _ {11}>J _ 1$ indicates a stronger intra-cell hopping compared to inter-cell hopping. Consequently, electrons favor forming dimer-like states. Furthermore, as $J_{11}>J_{33}$, next neighbour hopping does not disrupt the dimer state. Consequently, this explains the sequential emergence of dimers observed in Fig.~\ref{fig5}(c). Additionally, the Rényi entanglement entropy between the two outermost dimers and the remaining dimers remains near zero, as depicted in the left panel of Fig.~\ref{fig5}(f). When $\Delta_{J_{33}-J_{11}}>0$, the $2_{nd},\ (N-1)_{th}$ sites become topological edge states. This scenario, illustrated in Fig.~\ref{fig5}(a), exhibits long-range spin correlation, and the Rényi entanglement entropy between $2_{nd},\ (N-1)_{th}$ sites and other sites approaches zero as shown in Fig.~\ref{fig5}(d). 

\subsection{Magnetic-field-induced transition at quarter-filling}

In contrast to the isolated edge states observed at half-filling, the system at quarter-filling exhibits long-range antiferromagnetic (AFM) order across all topological phases. For the system parameters corresponding to Fig.~\ref{fig6}(a) and \ref{fig6}(b), the sites where the edge states appear are $(1 _ {st}, N _ {th})$ sites and $(1 _ {st}, 3 _ {rd}, (N - 2) _ {th}, N _ {th})$ sites, the long-range spin correlation characteristic of half-filling are absent at quarter-filling. Instead, a short-range AFM order, characterized by adjacent or sub-adjacent spin correlation, is observed. This short-range order, arising due to the existence of AFM order within the bulk, leads to a strong spin correlation between the edge states and the bulk. Consequently, the Rényi entanglement entropy does not exhibit a significant drop to zero as the $\Delta_J$ is changed, as illustrated in Fig.~\ref{fig6}(c) and \ref{fig6}(d) along the x-axis where $E_B=0$. 

The Hamiltonian describing the system under a magnetic field is given by $H_1=H-(E_B/2)\Sigma_i(\hat{n}_{i\uparrow}^s-\hat{n}_{i\downarrow}^s)$ with $E_B=g\mu_BB$, where $\mu_B$ is the Bohr magneton and $g$ is the g-factor of the electron in the material. According to the above discussion, the system exhibits no long-range spin correlation at quarter-filling. However, a sufficiently strong magnetic field can induce a transition to a maximally ferromagnetic ground state, as evidenced by the emergence of long-range spin correlation in Fig.~\ref{fig6}(c) and \ref{fig6}(d). This can be attributed to the magnetic field effectively transforming the quarter-filling state into a quasi half-filling state at high field strengths, which is similar to the true half-filling under zero magnetic field. Additionally, the Rényi entropy drops to zero for $\Delta_J>0$ due to the formation of  new topological edge states. 

The contours depicted in Fig.~\ref{fig6}(c) and \ref{fig6}(d) separate regions with distinct total spin $S_z$. For a one-dimensional chain with $N=8$, a maximally ferromagnetic ground state with four spin-up electrons at quarter-filling corresponds to $S_z=2$. Notably, the contour for $S_z=0$ disappears in Fig.~\ref{fig6}(c). This is a consequence of the degeneracy between $S _ z=1$ and $S _ z=0$ under zero field ($\omega=8$). When a magnetic field, regardless of its strength, is applied, $S _ z=1$ becomes the ground state of the system, leading to the disappearance of the $S_z=0$ contour. 

To achieve a maximally ferromagnetic state, the last spin-down electron needs to be pumped to the next unoccupied spin-up single-particle state. The existence of Hubbard $U$ causes that at quarter-filling, the energy of only one spin-down electron is higher than $U=0$ case. Consequently, a smaller critical magnetic field is required to overcome the energy barrier for flipping this spin-down electron to a spin-up state. Since it is very difficult to realize a strong magnetic field in experiment, the existence of Hubbard $U$ allows us to observe the above transformation under a much smaller magnetic field, as demonstrated in Fig.~\ref{fig6}(e) and \ref{fig6}(f).

Fig.~\ref{fig6}(e) suggests a potential demarcation line for $\Delta_{J_3-J_1}<0$. We infer that this is likely the result of a competition between the energy-lowering hopping term and the second-order process under the large U limit. However, a complete understanding of the underlying mechanism remains elusive. Further clarification regarding the demarcation line can be found in Appendix \ref{appendixA}. 

Although there will be a sign problem at quarter-filling, we can calculate the above results under different electron numbers using CP-AFQMC for a larger $1D$ chain with $N=28$, as shown in Fig.~\ref{fig7}. The presence of a magnetic field introduces topological edge states component to the ground state at quarter-filling. However, the trial wave function at quarter-filling obtained from a single-particle model does not contain edge component. Therefore, we can no longer simply use the single-particle wave function as the trial wave function. Here, we propose a new trial wave function by linearly superimposing the single-particle wave function at quarter-filling with the topological edge states. This new trial wave function at quarter-filling can be used to reduce the system's bias. 

In Fig.~\ref{fig7}(a)-\ref{fig7}(d), we employ the formula (\ref{spin corr}) to calculate the spin correlation. The results consistent with Fig.~\ref{fig6}, that is, the long-range AFM order manifested under the critical magnetic field and the long-range spin correlation above the critical magnetic field. Subsequently, the Rényi entanglement entropy of the system is calculated using formulas (\ref{Rényi}), as depicted in Fig.~\ref{fig7}(e). We can observe that for $\Delta_{J_3-J_1}<0$, the entanglement entropy remains nonzero regardless of the applied magnetic field . Conversely, when $\Delta_{J_3-J_1}>0$, the formation of long-range spin correlation leads to a drop in entanglement entropy towards zero as the magnetic field strength increases. This behavior is consistent with Fig.~\ref{fig6}(c) and \ref{fig6}(d), the entanglement entropy within each contour is a constant value. The calculated entanglement entropy for different spin configurations is connected by a polyline fit. The critical magnetic field, corresponding to the formation of the maximally ferromagnetic state, is defined as the magnetic field strength at which the entanglement entropy reaches its minimum for different values of $\Delta_{J_3-J_1}$. As illustrated in Fig.~\ref{fig7}(f),  the relationship between the critical magnetic field and the Hubbard $U$ is determined using CP-AFQMC. Similarly, a substantial decrease in the critical field with increasing Hubbard $U$ is observed. The case $J _ 1=J _ 3=0$ is detailed in the Appendix \ref{appendixD}. 

\section{SUMMARY AND DISCUSSION}\label{se4}

Topological edge states are a fundamental concept in topological physics, and their characterization is crucial for understanding the properties of topological materials. Spin correlation and entanglement entropy serve as essential measures for evaluating these edge states, even in strongly correlated systems. 

In this work, we employ two computational techniques, exact diagonalization and CP-AFQMC, to simulate the SSHH model incorporating next-nearest neighbour hopping. Through exact diagonalization calculations, we observed that in small systems with Hubbard $U$, spin correlation and Rényi entanglement entropy exhibit variations in response to the relative magnitude of system hopping. To extrapolate to the thermodynamic limit, we utilize the CP-AFQMC algorithm and employ the method outlined in \cite{PhysRevLett.111.130402} to calculate the system's Rényi entanglement entropy. The results obtained align with those from exact diagonalization but exhibit reduced finite-size effects. 

When the system is at quarter-filling, long-range AFM order emerges, leading to the disappearance of long-range spin correlation and a significant increase of the entanglement entropy. Interestingly, if an external magnetic field is applied at this case, the system reverts to a quasi-half-filling state, reintroducing long-range spin correlation and reducing entanglement entropy to near zero.

During the simulation of quarter-filling, the presence of a sign problem necessitates the employment of the CP-AFQMC method to mitigate the issue. In this process, we utilize modified trial wave functions to achieve enhanced convergence. By calculating spin correlation and entanglement entropy using CP-AFQMC, we can extrapolate the results to other interacting topological systems, obtaining findings that closely approximate the thermodynamic limit.

\begin{acknowledgments}
The authors would like to thanks Yuhang Lu and Pengyu Wen for help on computing resources. The authors also would like to thanks Wenjia Rao, Shuai Chen, and Zhen Guo for their helpful discussion. This work is supported by the National Natural Science Foundation of China under Grants No. 11974205, and No. 61727801, the Key Research and Development Program of Guangdong province (2018B030325002), and National Natural Science Foundation of China under Grant 62131002. 
\end{acknowledgments}

\appendix

\section{Comparison of ED and CP-AFQMC results with 1D $N=8$ chain and a puzzle}\label{appendixA}

\begin{figure}[H]
    \centering
    \includegraphics[width=0.4\textwidth]{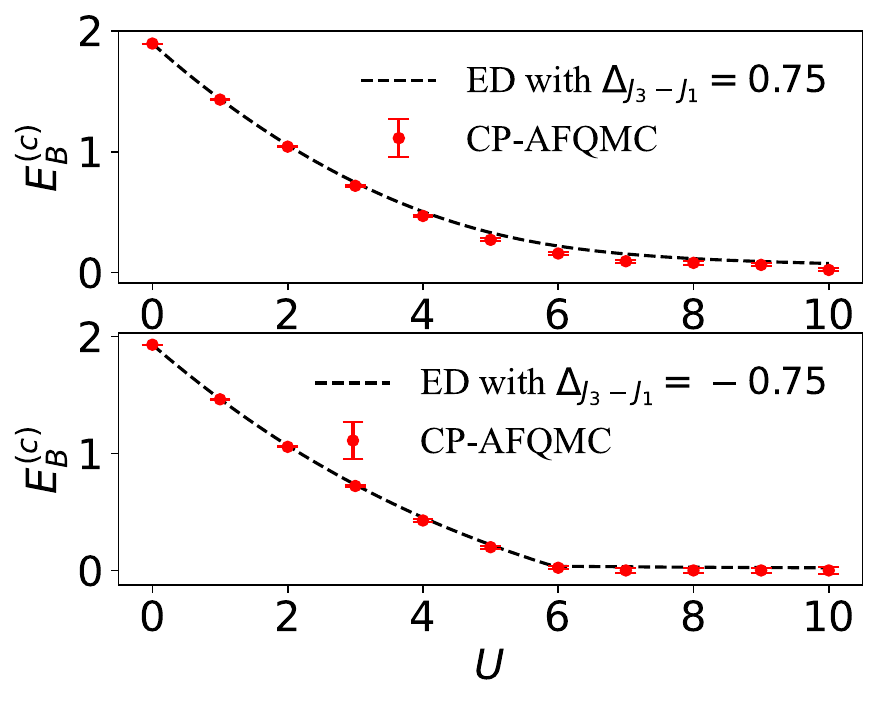}
    \caption{The magnitude of the critical field for a quarter-filling 1D chain with $N=8$ sites and electron numbers $n=4$ as a function of Hubbard $U$. }
    \label{fig8}
\end{figure}

To demonstrate the accuracy of CP-AFQMC results especially when there is a sign problem, we calculate the magnitude of the critical field for a quarter-filling 1D chain with $N=8$ sites as a function of Hubbard $U$, selecting $\Delta_{J_3-J_1}=-0.75$ and $\Delta_{J_3-J_1}=0.75$, which correspond to the leftmost and rightmost parameters in Fig.~\ref{fig6}(e). Due to the reduced system size, the imaginary time evolution length for CP-AFQMC calculations was appropriately shortened. This optimization improve computational efficiency without compromising accuracy for small systems, like the chain with $N=8$ sites and electron number $n=4$. The parameters we choose here are:
$\tau=0.01,\ N_{walker}=200,\ \beta=40\times 16\times \tau=6.4.$
Under the above parameters, we can calculate a relatively accurate value of the critical magnetic field, as shown in Fig.~\ref{fig8}. Compared with the exact diagonalization results, CP-AFQMC gives consistent calculation results. 

\begin{figure*}
     \centering
     \includegraphics[width=\textwidth]{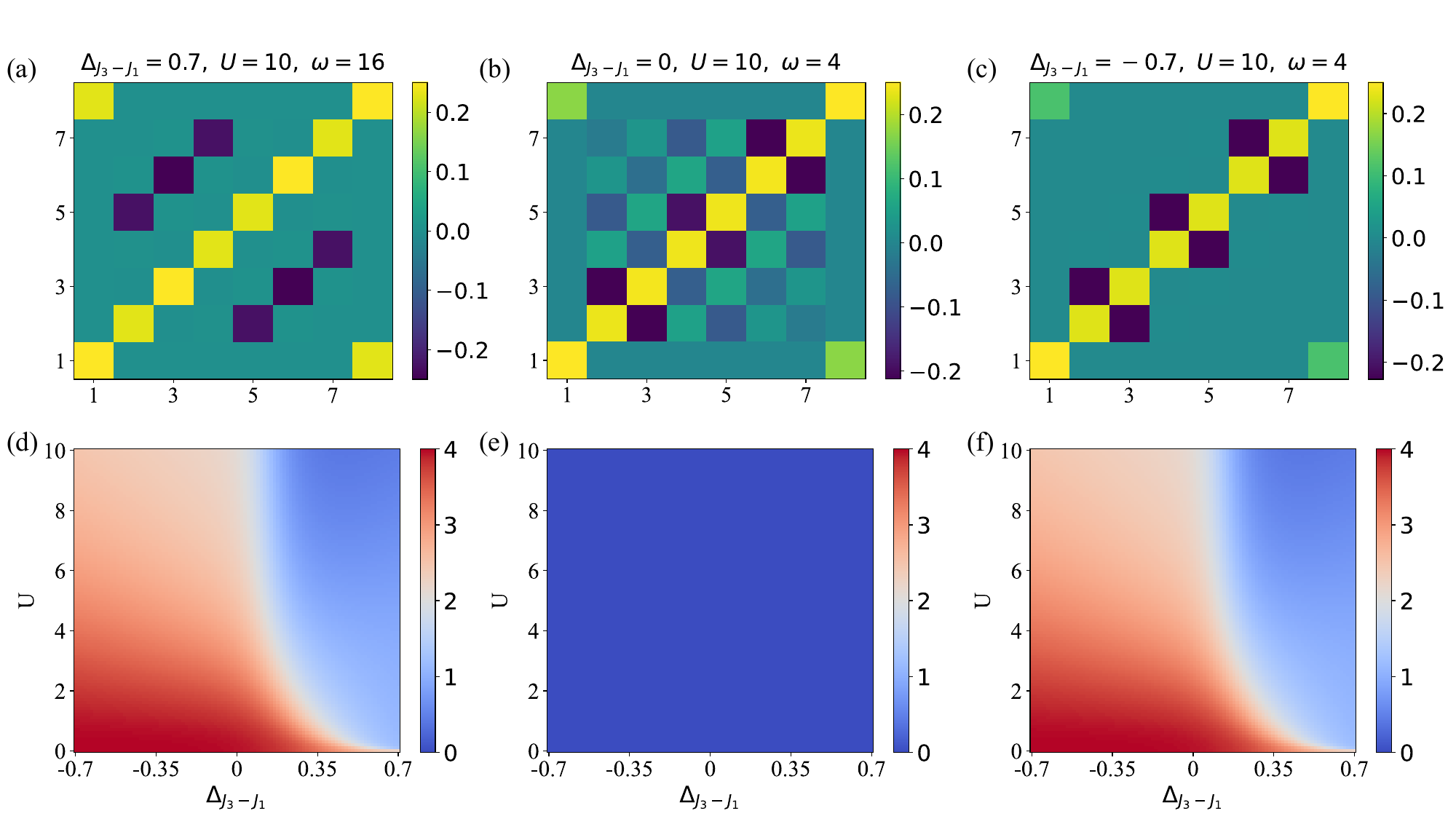}
     \caption{ED results same as Fig~\ref{fig4}. (a) Spin correlation $\langle S_{z,j}S_{z,k}\rangle$ with $\Delta_{J_3-J_1}=0.7$, topological edge states appear at [$1_{st}, 3_{rd}, 6_{th}, 8_{th}$] sites. (b) Same as (a) with $\Delta_{J_3-J_1}=0$ and there is a long-range anti-ferromagnetic order. (c) Same as (a) with $\Delta_{J_3-J_1}=-0.7$, and topological edge states appear at [$1_{st}, 8_{th}$] sites. And $\omega$ is the degeneracy of ground state. (d) The entanglement Rényi entropy between subsystem [$3_{rd}, 6_{th}$] sites and the remaining sites. (e) Same as (d) but subsystem [$1_{st}, 8_{th}$] sites and (f) with subsystem [$1_{st}, 3_{rd}, 6_{th}, 8_{th}$] sites. In (d)-(f), the position where the transition occurs is far from $\Delta _ {J _ 3 - J _ 1}=0$, which is result of finite-size effects. }
     \label{fig9}
\end{figure*}

\begin{figure*}
     \centering
     \includegraphics[width=\textwidth]{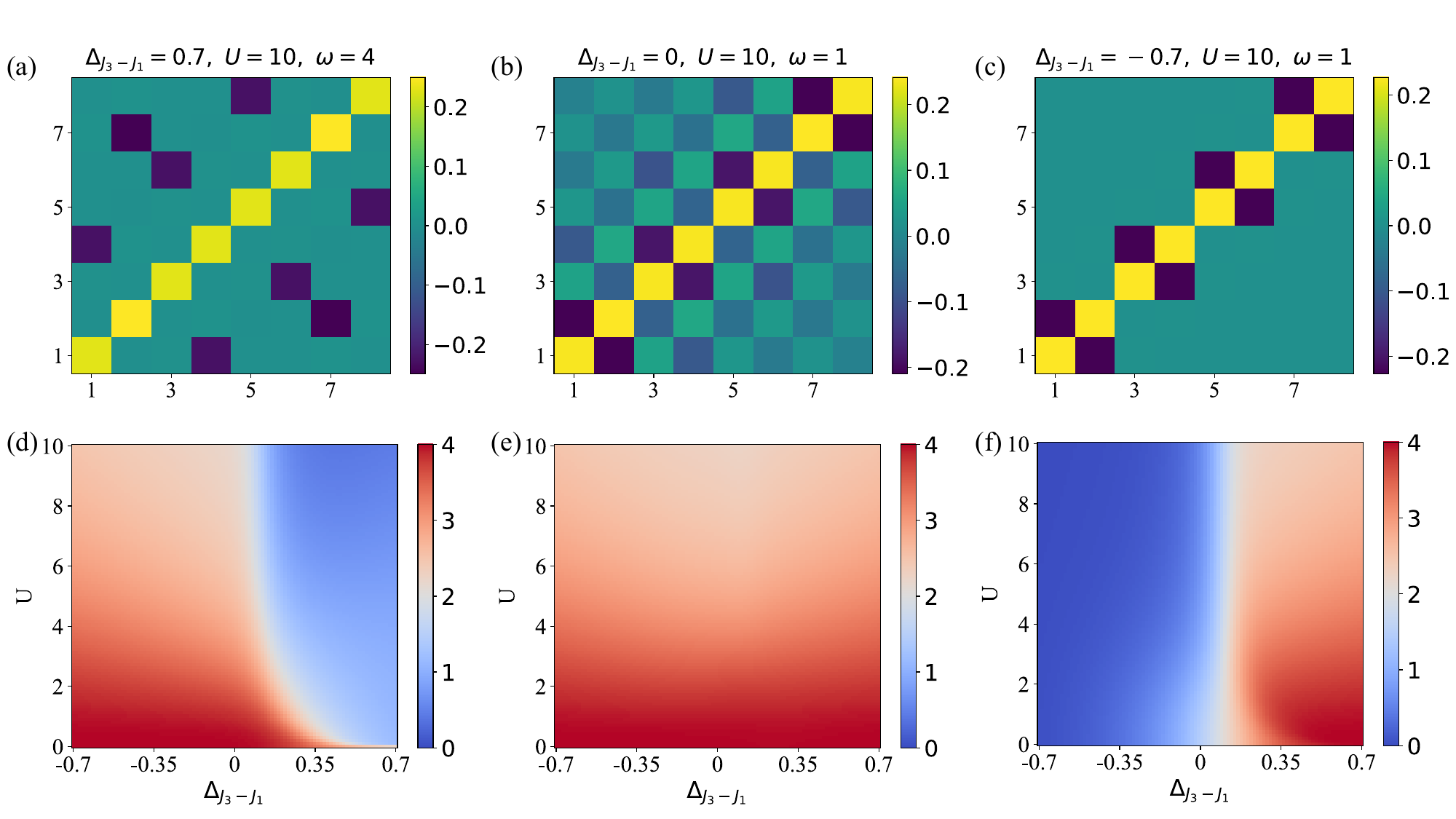}
     \caption{ED results same as Fig~\ref{fig5}. (a) Spin correlation $\langle S_{z,j}S_{z,k}\rangle$ with $\Delta_{J_{33}-J_{11}}=0.7$, topological edge states appear at [$2_{nd}, 7_{th}$] sites. (b) Same as (a) for $\Delta_{J_{33}-J_{11}}=0$ and there is a long-range anti-ferromagnetic order. (c) Same as (a) with $\Delta_{J_{33}-J_{11}}=-0.7$, and there are bulk dimers without topological edge states. (d) The entanglement Rényi entropy between subsystem [$2_{nd}, 7_{th}$] sites and the remaining sites. (e) Same as (d) but subsystem [$1_{st}, 8_{th}$] sites and (f) with subsystem [$1_{st}, 2_{nd}, 7_{th}, 8_{th}$] sites. And due to the formation of dimers, there will be a drop in Rényi entanglement entropy. }
     \label{fig10}
\end{figure*}

\begin{figure*}
     \centering
     \includegraphics[width=\textwidth]{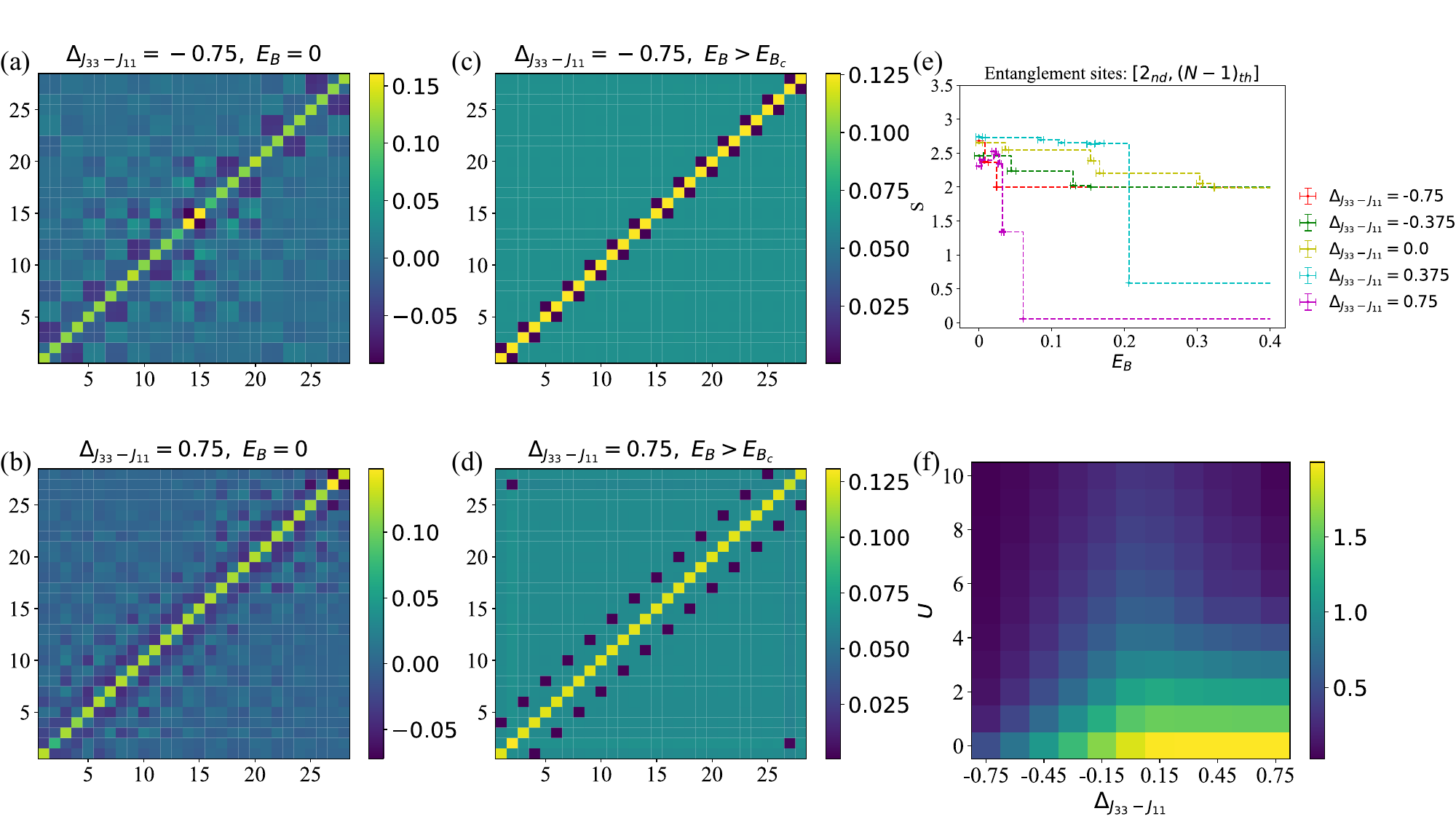}
     \caption{Same as Fig~\ref{fig7} for $J_{1}=J_{3}=0$, and $U=10$ for (a)-(e). (a) Spin correlation $\langle S_{z,j}S_{z,k}\rangle$ with $\Delta_{J_{33}-J_{11}}=-0.75$. (b) Same as (a) with $\Delta_{J_{33}-J_{11}}=0.75$ and topological edge states appear at [$2_{nd}, (N-1)_{th}$] sites. (c) Spin correlation $\langle S_{z,j}S_{z,k}\rangle$ with $\Delta_{J_{33}-J_{11}}=-0.75$ above the critical magnetic field. (d) Same as (c) but $\Delta_{J_{33}-J_{11}}=0.75$. (e) Rényi entanglement entropy with changing external magnetic field and hopping amplitude difference $\Delta_{J_{33}-J_{11}}$. (f) Critical field strength at various of the on-site interaction and hopping amplitude difference and it can be seen that for a certain $\Delta_{J_{33}-J_{11}}$, the value of the critical magnetic field will decrease as U increases. }
     \label{fig11}
\end{figure*}

The trend of the critical field change exhibits a clear segmentation effect in the figure below Fig.~\ref{fig8}. This is consistent with the demarcation line observed in Fig.~\ref{fig6}(e) as described in the main text. We speculate that this is caused by the competition between two processes, but we do not yet fully understand the underlying physical mechanism. Additionally, if we perform a vertical cross-section of Fig.~\ref{fig7}(f), we can also observe this phenomenon for $\Delta_{J_3-J_1}<0$. This indicates that the phenomenon is not caused by the finite-size effects of the system and requires further investigation. 

\section{Calculation of spin correlation in CP-AFQMC}
As before, we define the spin correlation as $C_{j, k} = \langle S_{z,j}S_{z,k}\rangle$ with $S_{z,j}=\frac{1}{2}(\hat{n}_{j,\uparrow}-\hat{n}_{j,\downarrow})$. And in CP-AFQMC we can calculate the spin correlation through the Green function

\begin{equation}
    \begin{aligned}
         C_{j,k} 
        & =\frac14\langle(\hat{n}_{j,\uparrow}-\hat{n}_{j,\downarrow})(\hat{n}_{k,\uparrow}-\hat{n}_{k,\downarrow})\rangle  \\
        &=\frac{1}{4}(G_{up}^{jj}G_{up}^{kk}-G_{up}^{kj}G_{up}^{jk}+G_{up}^{jk}\delta_{j,k}+G_{dn}^{jj}G_{dn}^{kk} \\ & \quad
        -G_{dn}^{kj}G_{dn}^{jk}+G_{dn}^{jk}\delta_{j,k} -G_{dn}^{jj}G_{up}^{kk}-G_{up}^{jj}G_{dn}^{kk}).
    \end{aligned}
    \label{spin corr}
\end{equation}
\label{appendixB}

\section{Spin correlation and entropy base on Exact Diagonalization}
\label{appendixC}

In this above study, we employ the CP-AFQMC method to compute the spin correlation and second-order Rényi entropy at half-filling. To validate the accuracy of the CP-AFQMC results, we perform additional calculations using exact diagonalization on a small system to obtain the long-range spin correlation and second-order Rényi entanglement entropy, as shown in Fig.~\ref{fig9} and Fig.~\ref{fig10}. The second-order Rényi entanglement entropy values obtained via ED exhibits a larger deviation from zero compared to the CP-AFQMC results. This discrepancy suggests a more pronounced finite-size effects in the calculations using ED, likely attributable to the inherent limitations of exact diagonalization when applied to small systems.

\section{quarter-filling with changing $\Delta_{J_{33}-J_{11}}$}
\label{appendixD}

To further substantiate our conclusions, we employ the CP-AFQMC method to investigate the behavior of the system under varying values of $\Delta_{J_{33}-J_{11}}$ as depicted in Fig.~\ref{fig11}. The results obtained corroborate the findings presented in Fig.~\ref{fig7}, demonstrating the restoration of long-range spin correlation and the disappearance of Rényi entanglement entropy above the critical magnetic field. Additionally, the critical field strength exhibits a gradual decrease with increasing values of the Hubbard $U$.

\clearpage
\bibliography{apssamp}

\end{document}